\documentclass[conference]{IEEEtran}
\usepackage{cite}

\usepackage{graphicx}
\usepackage{algorithmic}
\usepackage{algorithm}
\usepackage[fleqn]{amsmath}
\usepackage[tight,TABTOPCAP]{subfigure}

\begin{document}
%
\title{\textsf{WoT Store}: a Thing and Application Management Ecosystem for the W3C Web of Things}



%

\author{\IEEEauthorblockN{Luca Sciullo, Cristiano Aguzzi, Lorenzo Gigli, Luca Roffia, Angelo Trotta, Tullio Salmon Cinotti, Marco Di Felice}
\IEEEauthorblockA{\IEEEauthorrefmark{1} Department of Computer Science and Engineering, University of Bologna, Italy \\   \{luca.sciullo, cristiano.aguzzi, luca.roffia\}@unibo.it, lorenzo.gigli@studio.unibo.it\\ \{angelo.trotta5, tullio.salmoncinotti, marco.difelice3@unibo.it\}@unibo.it }
}


\maketitle

\begin{abstract}
In the next few years, the W3C Web of Things (WoT)  platform will represent a reference solution toward the deployment of fully interoperable systems, hence unlocking the potential of the IoT paradigm on several use-cases characterized by the current fragmentation of devices and technologies. At the same time, the worlwide adoption of the W3C WoT architecture depends on many factors, including also the availability of support tools that might facilitate the deployment of novel WoT applications or the integration with traditional IoT systems. To this purpose, the paper presents the \textsf{WoT Store}, a complete software platform enabling the discovery and management of W3C Things,  the monitoring of its properties and events, and the invoking of actions, all within the same dashboard. In addition, the platform leverages on the semantic description of each Thing with the goal of easing and automatizing the installation and execution of WoT applications, e.g. defining the behaviour of a Thing or implementing  mash-up operations from  multiple Things.  We sketch the main features, the architecture and the prototypal implementation of the  \textsf{WoT Store}. Moreover, we discuss the \textsf{WoT Store} capabilities on three IoT use-cases, i.e. industry 4.0, smart agriculture and home automation.

\end{abstract}

\section{Context}
Since the beginning, the Internet of Things (IoT) suffered from
lacks of  common interaction and communication paradigms that translated into the creation of several, independent technology silos, with severe concerns on the deployment of heterogeneous systems. The interoperability  can be considered both a challenge and a market opportunity: according to \cite{Summary15}, almost 40\% of the value of  IoT markets can be unlocked when achieving full interoperability among heterogeneous IoT systems. Among the several research projects and standardization initiatives proposed so far and addressing  the IoT interoperability issue, a key role is played by the World Wide Web Consortium (W3C), i.e. the leader of  Open Web platforms, and more specifically by the Web of Things (WoT) working group \cite{WoT}; this latter has been/is working on the  proposal of a platform-independent API and of service discovery mechanisms for the full IoT  interoperability. Clearly, a worldwide acceptance of the W3C WoT platform is fundamental for the purpose of mitigating the fragmentation of WoT-related initiatives, and of the definition of a reference solution;  on the opposite, there is always the risk of a new IoT technology silo. To this purpose, it is worth remarking that the software acceptance in industrial scenarios is  largely influenced by the size and the quality of its Software Ecosystem (SECO). This latter can be defined as the set of software solutions that enable, support, and automatize the activities and transactions by the actors in an associated social or business scenario. As reported in \cite{barbosa122011systematic}, several studies in the literature investigate the success of a software solution and its SECO  as vehicle to attract new players. The availability of supporting software modules is acknowledged as the key success of any platform in \cite{Iyer2007MonitoringPE}. Furthermore, the availability of a large SECO can translate into several other benefits, like the cooperation and the knowledge sharing among multiple and independent software vendors and the cost reduction in the software development \cite{SECObenifits}.\\
We believe that the presence of well-defined and complete support tools can become a turning point toward the large-scale adoption of W3C WoT solutions. This paper provides a significant contribution in this direction, by presenting the \textsf{WoT Store}, a software platform that ease the management of Things and the deployment of WoT applications on top. The \textsf{WoT Store} can be considered a component of the W3C WoT SECO, as well as an enabler of novel services and applications for the W3C WoT SECO, hence contributing to the further dissemination and utilization of the W3C WoT standards. We would like to highlight that the contribution presented in this study is the culmination of the research activities conducted by our research lab in the field of IoT interoperability platforms and the WoT. In 2017, we joint the W3C Face to Face meeting (Dusseldorf, Airport Conference Center) and discussed how to support the emerging WoT platforms via the Dynamic Linked Data and the SEPA technology \cite{SEPA}. This latter is a semantic broker enabling the detection and the notification of changes over Linked Data endpoints by means of content-based publish-subscribe mechanisms. The \textsf{WoT Store} exploits some functionalities of  the SEPA tool for the Thing discovery, while adding a suite of novel sotware modules for the Thing management, the WoT application management and the IoT data management.\\
The  paper is organized as follows. Section II describes the features,  architecture and implementation technologies of the \textsf{WoT Store}, while section III discusses its suitability on three different application domains, i.e. the Industry 4.0, Smart Agriculture and Home Automation. Finally, open issues and current  extensions are discussed in  Section IV.

\section{The \textsf{WoT Store}}
\label{wot}
As previously mentioned, the W3C Web of Things (WoT) standards constitute a novel and effective approach to enable the seamless integration of heterogeneous IoT components, abstracting from their implementation \cite{WoT}. Several IoT markets, traditionally characterized by the fragmentation of the protocols and communication technologies, might greatly benefit from the definition of a reference architecture providing a way to describe, in a non-ambiguous way, the interfaces and the interaction patterns of the IoT components \cite{Ji18}\cite{Klotz18}\cite{McCool18}\cite{Blank18}. At the same time, the goal of our research was not to target a novel application scenario of the W3C WoT, rather to leverage on the novel opportunities offered by a world of W3C Web Things, being them native (i.e.  implementing the architecture defined in \cite{WoT} from scratch) or adapted from traditional deployments (e.g. via a proxy). More specifically, we addressed the following key questions: (\textit{i}) how to ease the discovery and the management of W3C Things, by supporting both private (i.e. only visible within a local network environment) and public (i.e. reachable by the whole Internet) environments? The WoT management means general-purpose functionalities like for instance: find the Things satisfying specific requirements (e.g. location), perform  actions on them, display property values, etc that we expect to be present in any W3C WoT deployment, regardless of the use-case. (\textit{ii}) given that the Thing Descriptor (TD) provides a computer-readable interface of the behaviour of a Thing and of the way to interact with it, how can we ease and even automatize the deployment and  execution on  WoT applications composed by multiple, heterogeneous Things?\\ 
We addressed  the  questions above in the \textsf{WoT Store}, a generic software platform for the management of W3C-compliant Things and applications of the WoT SECO. A prototype version of the \textsf{WoT Store} has been presented in \cite{Sciullo19}.  The overall framework is designed to be highly modular; different modules can be loaded/unloaded or customized for specific application use-cases, as we  discuss further in Section \ref{usecases}. We present the main modules in the Section below, while the architectural and implementation details are sketched in Section \ref{architecture} and \ref{implementation}, respectively.


\subsection{Features}
\label{wotstore_sec}
The features offered by the \textsf{WoT Store} platform can be grouped in three main modules:

\subsubsection{Things Manager}
The first module allow users to handle and manage all the Things deployed in the environment.  It works as a dashboard, i.e. users can visualize and monitor the Things’ status and properties. Moreover, it provides basic mechanisms to interact with the Things, by invoking, for instance,  actions with custom inputs. Finally, it integrates discovery mechanisms so that new Things can dynamically be loaded on the \textsf{WoT Store} as soon as they are deployed.
\subsubsection{Applications manager}
The second module supports the installation and execution of IoT Applications, by considering two classes of software, i.e.: Things Applications (TA) and Mash-up Applications (MA), as better detailed in \cite{Sciullo19}. The \textsf{WoT Store} mimics the operations of a mobile application markets, since users can download and install the applications on their Things with minimal manual efforts; in addition, there are sharing mechanisms that allow developers to upload the code, revise it and leave line comments. Each application is  provided with a semantic description, hence  users can issue semantic queries in order to find the perfect match between their needs, the available Things exposed by the Thing Manager and the applications available on the store. Finally, the module  offers the possibility to run MAs directly on the cloud, so that no computational resources are requested on the client side. 
\subsubsection{Data manager}
The third module includes functionalities for the  visualization, processing and manipulation of the data produced by the deployed Things. More in details, the \textsf{WoT Store} provides data filtering mechanisms  according to the data origin, i.e, the producer Things, and additional components to plot the data on the dashboard. Different data flows can be aggregated by installing the proper MA. We plan to include  data-analytics sub-modules as future work. 
\subsection{Architecture} 
\label{architecture}
\begin{figure} 
\centering
\includegraphics[scale=0.3]{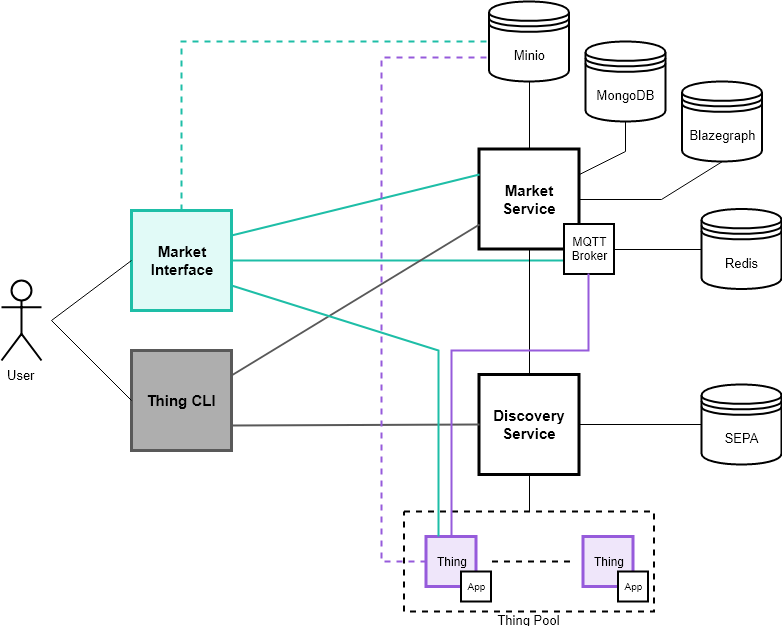}
\caption {The \textsf{WoT Store} Architecture}
\label{fig:wots_architecture}
\end{figure}

The overall architecture of the \textsf{WoT Store} is shown in Figure \ref{fig:wots_architecture}. We briefly introduce each component in the following.
On the left part of the Figure there are the interfaces allowing the user to interact with the platform:
\begin{itemize}
    \item \textit{Market Interface} (MI): this consists of a Web application offering the main features of the store previously listed in Section \ref{wotstore_sec}. The component communicates directly with: the Market Service (MS) for all the resource operations and the real-time notifications; the available Things (rendering them and displaying the property values); the MQTT Broker, which is used as collector of messages related to events triggered by the Things.
    \item \textit{Thing CLI} (TC): this is constituted by a command line tool that helps the configuration and then the online publication of a Thing, or of a group of Things. The users are guided during the deployment of their application scripts and the definition of  the semantic metadata. At the end of the process, each new Thing is registered on the Discovery Service and on the Market Service.
\end{itemize}
The back-end of the \textsf{WoT Store} platform consists of two main APIs services:
\begin{itemize}
    \item \textit{Market Service} (MS): this consists of a REST API enabling the interaction with the main system resources, such as the Applications, Things, Users and so on. CRUD operations are implemented for each category of resource (and several other resource-specific calls). It also exposes WebSockets for real-time data retrieval and updates.
    \item \textit{Discovery Service} (DS): this service monitors the status of each Thing. Whenever a change occurs (voluntary or not), the DS keeps track of the new state, then notifies the MS which updates the Thing Description (TD) correspondingly.
\end{itemize}

\subsection{Implementation} 
\label{implementation}
The MS has been implemented as a \texttt{Node.js} v10.x application using the \texttt{LoopBack} v3 framework and a MQTT Broker built with the \texttt{Mosca} library. The current deployment includes four types of databases/storage:
\begin{itemize}
    \item \texttt{Minio}: this is an object storage server used to store the applications source code.
    \item \texttt{MongoDB}: this database contains user data, relationships between resources and other utility models.
    \item \texttt{Blazegraph}: this triplestore contains the application metadata and the TDs.
    \item \texttt{Redis}: this is a high performance in-memory database, required for the real-time processing of MQTT messages.
\end{itemize}
The MI has been implemented as a an \texttt{Angular} v6.x Web application composed of several  modules. Among the various libraries, we find \texttt{@angular/material} for quality Material Design elements, \texttt{ngx-mqtt} and \texttt{socket.io-client} for communication.
TC is another Node.js application that uses packages like \texttt{chalk},  \texttt{clui} and  \texttt{inquirer} to build the interactive terminal; \texttt{request} and   \texttt{shelljs} modules are employed in order to manage the HTTP calls.\\
The \textsf{WoT Store} assumes a Servient to be installed on devices. Currently the only W3C-supported and guaranteed implementation of the Servient  is the one  by the Eclipse Foundation  (Eclipse \texttt{Thingweb node-wot} \cite{servient}).
The Node.js implementation requires some additional dependencies to be executed: \texttt{Python} 2.7, \texttt{make} and a \texttt{C/C++} compiler. Our TC downloads the Servient directly from the repository, configures it according to user preferences and runs it by passing the application scripts as arguments.

\begin{figure} 
\centering
\includegraphics[scale=0.09]{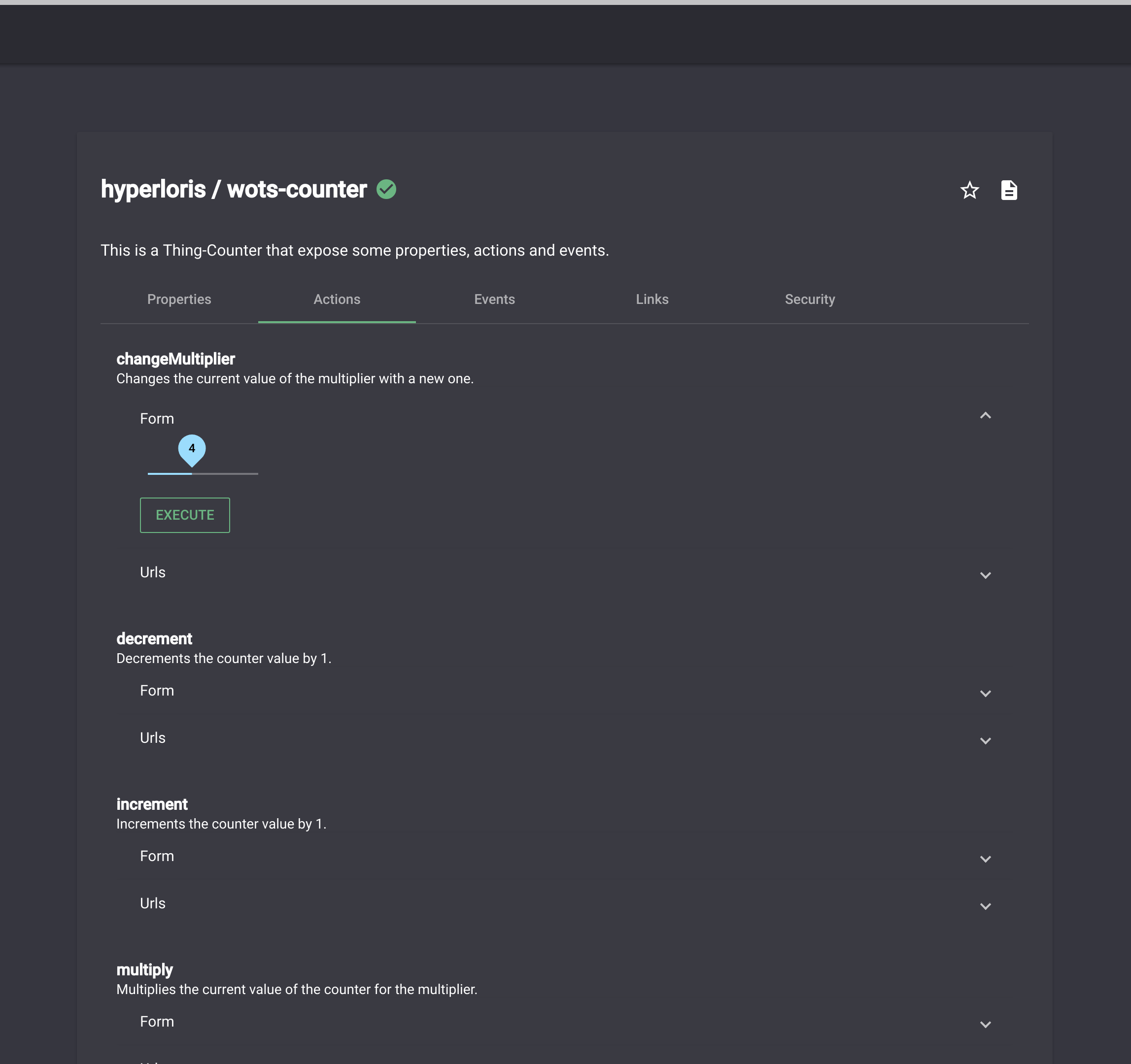}
\caption {The \textsf{WoT Store} MI - view of a Thing (actions)}
\label{fig:wots_dashboard}
\end{figure}

\section{Use Cases}
\label{usecases}
The \textsf{WoT Store} can be considered  a component and an enabler of the W3C WoT SECO; the main goal is to ease the deployment of heterogeneous systems, with advantages in terms of time and costs reduction. The application scenarios of the  \textsf{WoT Store} are numerous, and comprises several IoT markets active so far. We discuss three of them,  with explicit references to the research projects which see the participation of the authors \cite{swamp}. 
\subsection{Industry 4.0}
The \textsf{WoT Store}  supports the management of large number of heterogeneous devices via a common dashboard. For this reason, one of the main application scenario is constituted by  Industry 4.0 environments characterized by the  large-scale installation of sensors, and by the need to process vast amount of sensor data in order to build the digital twin of a physical component. We can assume that each sensor is represented by a Thing and runs a default factory application that publishes itself on the DS. Once Things have been deployed, it is possible to display them on the main console of the \textsf{WoT Store} thanks to the Discovery module, and take advantage of the functionalities provided by our tool, like for instance:  \begin{itemize}
    \item \textit{Update-all}: this feature allows to replace the  TA currently deployed on a set of Things with a single action. Let us imagine a situation where a bug is discovered on the current TA, and hence a patch must be applied on all the Things. Without the  \textsf{WoT Store}, this would require a manual  re-configuration of all the devices, with a clear impact in terms of time and cost.  The  \textsf{WoT Store}  includes a remote, secure mechanism to dynamically change the  TA code currently executed on a Thing; moreover, the users can issue a semantic query in order to update a set of Things all satisfying the same query conditions (e.g. "update all sensor devices of type temperature available in room 262").
    \item \textit{Mashup Application}: after Things have started  their TAs, data should be collected and aggregated according to  user-defined policies. For this reason, directly from the console of \textsf{WoT Store}, technicians can write and run a custom MA for Things querying and data processing. 
    \item \textit{Data analysis}:  the \textsf{WoT Store} dashboard includes  built-in mechanisms for plotting the data produced by set of Things and analyzing the aggregated results. 
\end{itemize}

\subsection{Smart Agriculture}
Smart Agriculture leverages on IoT solutions to implement the so called  Third Green Revolution. In this context, ICT technologies face unique challenges like, among others: the lack of stable power supply, the need of calibration procedures customized for every type of soil, the security concerns from farmers, just to name a few \cite{swamp}.  The role of the \textsf{WoT Store} is similar to the Industry 4.0 use case, since our platform can be considered  an open ecosystem where different stakeholders can interact to effectually collect and process data from crops.  
Let us consider for example the digitization of wheat production in farms. Commonly, various sensors/actuators are deployed to enhance the productivity and the economical gain of the farm: soil moisture sensors, drones, automatic sprinklers and cameras. The deployment of the technological infrastructure consists in the major effort from both technicians and farmers. Moreover, in most of the crop types, the process should be repeated every year. In this scenario, the Thing Manager of the \textsf{WoT store} can be used to ease the installation process and automatize most of the configuration phases. Furthermore, using the data flow analysis, agronomists or soil experts can create a calibration models for soil moisture sensors to better fit the real volumetric water content. Later, the model could be uploaded to the system as a MA or the Thing Manager can be used to  modify directly the sensor parameters (i.e. invoking an action with the new calibration curve or changing a conversion parameter). The open platform can be also exploited by IT companies specialized in data mining techniques, IA or agronomic models. Such software applications can be offered as MA applications and installed in a farm system with a simple add on. 
Other functionalities of  the \textsf{WoT Store} can be used, similarly to the  previous case: 
\begin{itemize}
    \item \textit{Update-all}:  Industry farmers need to keep TA up-to-date in order to avoid data losses and discontinuous services.
    \item \textit{MA deployment}: customization is the key element of every Smart Agriculture application, since   crops are not always the same and several variables should be taken in to account, like soil, weather, geographical position, irrigation and cultivation methods. MAs can be deployed and instantiated hence  helping the customization process even by non ICT experts.
    \item \textit{Data analysis}: Using data analysis tools embedded in the platform we can support the decision makers in the customization process described above.   
\end{itemize}
\subsection{Home automation}
The current IoT market for home automation   is characterized by the proliferation of devices mapped on different hardware/software technologies,  often adopting  proprietary solutions. This has lead to the emergence of  the  technology silos previously mentioned. Thanks to the W3C WoT, different silos might be integrated by adding a semantic description to each interface, hence without requiring changes to the original deployments. Again, Things can  be displayed and managed through the     \textsf{WoT Store}, which can provide the following benefits in this use-case: 
\begin{itemize}
    \item \textit{New Thing integration}: 
    when users need to add a new device to their home automation system, they can take advantage of the self-publishing service of Things on the \textsf{WoT Store}, if already W3C WoT compliant. Otherwise, through the \texttt{Thing CLI} they can easily set-up and configure a new Thing for the current device.
    \item \textit{User Interface}:  the \textsf{WoT Store} provides an easy and customizable configuration panel through which users can monitor the behaviour of their home devices.
    \item \textit{Mashup Application}: after having configured their Things, even not-technical users can set-up their MAs providing user-defined automatized actions.
\end{itemize}

\section{Conclusion and Future Works}
\label{conclusion}
In this study we presented the \textsf{WoT Store}, a novel platform for the management of W3C-compliant Things and Applications. We  presented a preliminary version in \cite{Sciullo19}; the platform is still under deployment since there are plenty of additional features that can be integrated. Among the others, we mention two current works. First, we plan to extend the platform with the integration of data analytics features, based on machine learning and data mining techniques. Second, we are working on a control access mechanism in order to manage all the access requests made by and toward each Things directly from the \textsf{WoT Store}.

\end{document}